\documentclass[a4paper,fleqn,usenatbib]{mnras}
\usepackage{newtxtext,newtxmath}
\usepackage[T1]{fontenc}
\usepackage{ae,aecompl}

\usepackage{graphicx}   % Including figure files
\usepackage{amsmath}    % Advanced maths commands
\usepackage{amssymb}    % Extra maths symbols

\title[QPPs with multiple periods]{Quasi-periodic pulsations with multiple periods in hard X-ray emission}

\author[D. Li $\&$ Q. M. Zhang]{
D. Li,$^{1,2}$\thanks{E-mail:lidong@pmo.ac.cn} and Q. M. Zhang$^{1}$ \\
$^{1}$Key Laboratory for Dark Matter and Space Science, Purple
Mountain Observatory, CAS, Nanjing 210008, PR China \\
$^{2}$Key Laboratory of Modern Astronomy and Astrophysics (Nanjing
University), Ministry of Education, Nanjing 210023, PR China}

\date{Accepted XXX. Received YYY; in original form ZZZ}

\pubyear{2017}

\begin{document}
\label{firstpage}
\pagerange{\pageref{firstpage}--\pageref{lastpage}}
\maketitle

\begin{abstract}
We explore the quasi-periodic pulsations (QPPs) with multiple
periods in hard X-ray (HXR) emission from {\it Fermi}/GBM during the
impulsive phase of solar flare (SOL2014-09-10). The completely new
observational result is that the shorter periods appear at lower
energies of the X-ray photons at the beginning and the longer
periods appear at higher energies at the end, with some intersection
of the periods at medium energies. We also find the shorter and then
the longer periods during the same phase of this flare. Using the
wavelet power spectrum and fast Fourier transform (FFT) spectrum, we
analyze the normalized rapidly varying signal divided by its slowly
varying signal, which is the smoothed original HXR flux. The periods
of 27~s and 37~s are derived at lower energy channels between
17:25~UT and 17:29~UT, i.e., 12.0$-$27.3~keV and 27.3$-$50.9~keV.
Then the periods of 27~s, 46~s and 60~s are observed at
medium-energy channel from 17:26~UT to 17:33~UT, such as
50.9$-$102.3~keV. And the period of 80~s is detected at higher
energy channel from 17:28~UT to 17:33~UT, such as 102.3$-$296.4~keV.
\end{abstract}

\begin{keywords}
Sun: flares -- Sun: oscillations -- Sun: X-rays, gamma-rays

\end{keywords}

%%%%%%%%%%%%%%%%% BODY OF PAPER %%%%%%%%%%%%%%%%%%

\section{Introduction}
Quasi-periodic pulsations (QPPs) are usually observed in the light
curves of solar flares
\citep{Nakariakov09,Inglis16,Pugh16,Van16,Zhang16a}. They typically
display regular and periodic peaks from the total flux based on the
time-series analysis. The observations of QPPs can cover almost all
the wavelengths, such as soft/hard X-rays (SXR/HXR)
\citep{Lipa78,Ning14a,Aschwanden98,Dolla12,Tan16},
extreme-ultraviolet (EUV/UV)
\citep{Nakariakov99,Ning14b,Liu11,Kumar16,Lil16} and radio
\citep{Aschwanden94,Ning05,Tan10,Yu13,Kupriyanova16} emission. In
addition, QPPs can be detected with spectral observations, i.e.,
Doppler shifts, line widths and intensities
\citep[e.g.,][]{Ofman02,Wang02,Tian11,Brosius15,Li15,Tian16}.

So far, it is still unclear which mechanism could cause the QPPs
\citep{Aschwanden94,Nakariakov09,Van16}. Previous findings show that
QPPs could be interpreted as the MHD wave
\citep{Roberts84,Nakariakov05,Nakariakov09,Tian16} or produced by
the quasi-periodic magnetic reconnection
\citep{Karlicky05,Nakariakov06,Liu11,Li15,Li17a}. These QPPs models
only apply to one or several events, but fail to explain all the
observational features.

The observed periods of QPPs range from tens of seconds to tens of
minutes
\citep{Lipa78,Aschwanden94,Karlicky05,Tan10,Shen12,Shen13,Tan16,Tian16,Ning17}.
Sometimes, QPPs in the same event could show multiple periods in
radio \citep{Inglis09}, H$\alpha$ \citep{Srivastava08,Yang16}, EUV
\citep{Su12}, SXR \citep{Chowdhury15}, and HXR \citep{Zimovets10}
bands. The period ratio of QPPs in the same event depends on the
magnetohydrodynamic (MHD) mode \citep{Nakariakov05}. The ratio
equals or almost equals two for the weakly dispersive MHD modes,
i.e., torsion, slow magneto-acoustic, and kink. However, it deviates
from two for the highly dispersive modes. The difference the period
ratio from two is also attributed to additional physical effects
like siphon flows \citep{Li13}, the loop expansions \citep{Verth08},
or the longitudinal density stratifications \citep{Andries05}.

QPPs with multiple periods can be detected during all phases of
solar flare, i.e., from impulsive to decay phases
\citep{Reznikova11,Wang12,Hayes16,Tian16}. However, the fact that
the dominant period of flare QPPs becomes longer with the increasing
photon energy during the impulsive phase has not been reported yet.
Using the high temporal resolution observations from {\it Fermi}
Gamma-ray Burst Monitor (GBM) \citep{Meegan09}, we explore the QPPs
behaviors in a solar flare on 2014 September 10. The new
observational finding may be a challenge to the QPPs theory, and
could help to refine the model of flare QPPs.

\section{Observations}
\begin{figure}
\includegraphics[width=\columnwidth]{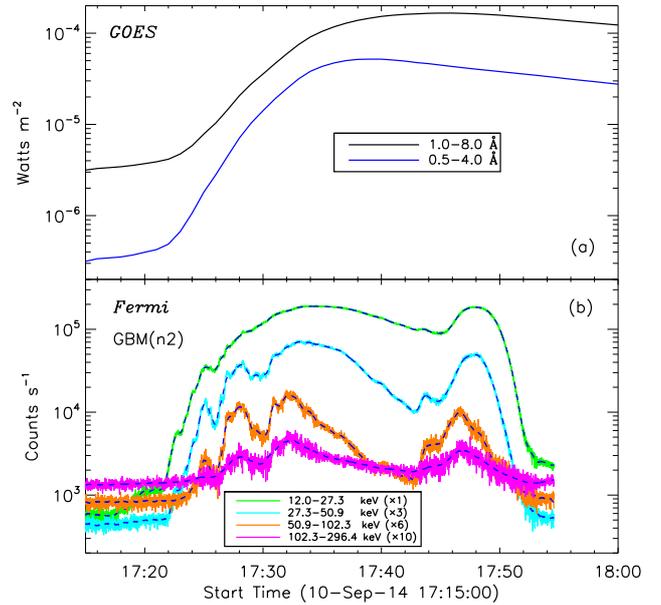}
\caption{Panel~(a): {\it GOES} SXR flux from 2014 September 10 flare
at 1.0$-8.0$~{\AA} (black) and 0.5$-$4.0~{\AA} (blue). Panel~(b):
{\it Fermi} HXR light curves at 12.0$-$27.3~keV (green),
27.3$-$50.9~keV (turquoise), 50.9$-$102.3~keV (orange), and
102.3$-$296.4~keV (purple). The blue dashed profiles represent the
slowly varying signals from original HXR light curves. Each HXR
light curve is multiplying by a factor.} \label{flux}
\end{figure}

In this paper, we investigate an X1.6 flare which takes place in
NOAA AR 12158 on 2014 September 10. It is accompanied by a fast
coronal mass ejection (CME) \citep{Cheng16}. It starts at about
17:21 UT, and then reaches its maximum at around 17:45 UT from the
{\it GOES} SXR flux at 1.0$-$8.0~{\AA} and 0.5$-$4.0~{\AA},
respectively, as shown in Figure~\ref{flux}~(a).
Figure~\ref{flux}~(b) shows the light curves at four HXR channels
from {\it Fermi}/GBM, such as 12.0$-$27.3~keV, 27.3$-$50.9~keV,
50.9$-$102.3~keV, and 102.3$-$296.4~keV. To display the HXR light
curves clearly, we have multiplied a factor for every HXR light
curve. They are observed by n2 detector, whose direction angle to
the Sun is relative stable during the impulsive phase of solar
flare, i.e., between 17:21 UT and 17:45 UT. However, the other
detectors frequently change their direction angles. After the
impulsive phase (i.e., 17:45~UT), the n2 detector also shifts its
direction angle frequently, this results into the HXR peak around
17:47~UT, which may be not real. It is missing data after 17:54~UT.
The time cadence of {\it Fermi} light curves is about 0.256~s, but
it automatically becomes to 0.064~s when the flare bursts.
Therefore, the light curves from {\it Fermi} in
Figure~\ref{flux}~(b) are interpolated into an uniform cadence,
i.e., 0.256~s \citep[see.,][]{Li15,Ning17}.

\section{Data reduction technique}
To look closely the periods of QPPs, we decompose every light curve
into slowly varying signal and rapidly varying signal. The slowly
varying signal is obtained by smoothing (SMOOTH.PRO in IDL) the
original data with a timescale boxcar
\citep{Reznikova11,Dolla12,Li15,Hayes16}, as shown by the dashed
profile overlapping on the solid profile in Figure~\ref{flux}~(b).
Previous observation has been suggested that this X1.6 flare may
exhibit QPPs with short time periods in HXR bands, i.e., 30~s and
45~s \citep{Ning17}. However, these short timescale vibrations have
much smaller amplitude than the strong background emission, which
make them difficult to be detected \citep{Dolla12,Hayes16}.
Therefore, the window timescale of 30~s is chosen to highlight the
short timescale fluctuations. The rapidly varying signal is achieved
by subtracting the slowly varying signal from its original light
curve. And then the rapidly varying signal is normalized by its
slowly varying signal, so that it could exhibit the QPPs behaviors
much better \citep[see.,][]{Kupriyanova10,Kupriyanova13,Li17}.
Figure~\ref{qpp1}~(a) shows the normalized rapidly varying signal at
HXR 12.0$-$27.3 keV between 17:22~UT and 17:38~UT from {\it Fermi}
observations. It clearly shows the periodic peaks and could be
identified as QPPs. Note that we only select some time-series flux
during the impulsive phase to avoid the frequently changing of the
direction angle of n2 detector.

\section{Results of analysis}
\begin{figure}
\includegraphics[width=\columnwidth]{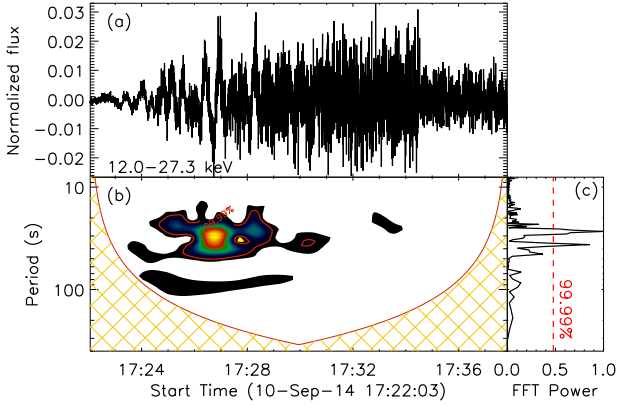}
\caption{Panel~(a): The normalized rapidly varying signal at
12.0$-$27.3 keV between 17:22~UT and 17:38~UT. Panel~(b): Wavelet
power spectrum. Panel~(c): FFT spectrum. The red contours and
vertical line outline the 99.99\% confidence levels.} \label{qpp1}
\end{figure}

\begin{figure}
\includegraphics[width=\columnwidth]{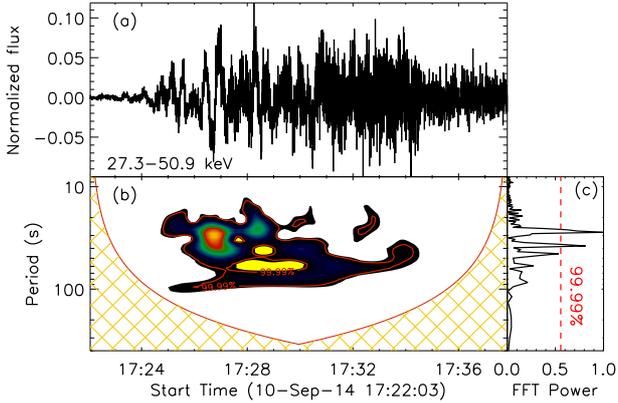}
\caption{Same as Figure~\ref{qpp1}, but at 27.3$-$50.9 keV.}
\label{qpp2}
\end{figure}

Figure~\ref{qpp1}~(b) shows the wavelet power spectrum from the
normalized rapidly varying signal at lower energy channel, i.e.,
12.0$-$27.3 keV, the red contours outline the significance levels of
99.99\%. It displays the dominant period with a broad band between
17:25~UT and 17:29~UT, as for most QPPs in microwave and HXR bands
\citep{Nakariakov09,Dolla12,Hayes16}. To further confirm the
periodicity, we perform a periodogram analysis for the normalized
rapidly varying signal with the Lomb-Scargle periodogram method
\citep{Scargle82}, as shown in panel~(c). The red vertical line
represents the 99.99\% confidence level defined by \cite{Horne86}.
It shows two peaks above the confidence level, and both them have
narrow widths. Therefore, the dominant period is obtained from the
peak value in the Fourier periodogram, and the error bars are
calculated from both the width of the spectral peak in the
periodogram and the discreetness of the Fourier frequency grid. Thus
two dominant periods within an error bar can be estimated from
17:25~UT to 17:29~UT at 12.0$-$27.3~keV, which are 27$\pm$2.8~s and
37$\pm$3.5~s. The same results can be found at 27.3$-$50.9~keV
during almost the same time intervals, as seen in Figure~\ref{qpp2}.
Although the wavelet spectrum still displays weak signal of
periodicity after 17:29~UT (panel~b), the FFT spectrum only shows
two peaks within narrow widths greater than the confidence level at
about 27~s and 37~s, which are similar to that in
Figure~\ref{qpp1}~(c). Thus we could derive the periods of
27$\pm$2.8~s and 37$\pm$3.5~s at 27.3$-$50.9~keV between 17:25~UT
and 17:29~UT.

\begin{figure}
\includegraphics[width=\columnwidth]{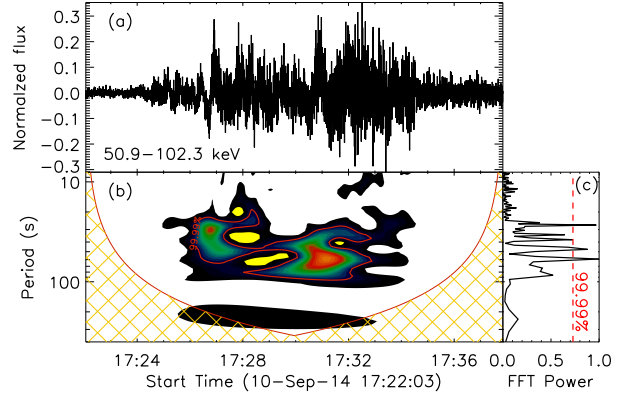}
\caption{Same as Figure~\ref{qpp1}, but at 50.9$-$102.3 keV.}
\label{qpp3}
\end{figure}

With the growth of HXR photon energies, the period of QPPs changes
obviously to the larger value, as shown in Figures~\ref{qpp3}. At
medium-energy channel (50.9$-$102.3~keV), the QPPs display multiple
periods within a broad band from 17:26~UT to 17:33~UT. The shorter
periods appear at the beginning of flare impulsive phase, such as
between 17:26~UT and 17:29~UT. And the longer period appears at the
end of flare impulsive phase, i.e., between 17:28~UT and 17:33~UT
(Figure~\ref{qpp3}~b). Considering the peaks within narrow width in
FFT spectrum, we can get three dominant periods within an error bar
at 50.9$-$102.3~keV, they are 27$\pm$1.8~s, 46$\pm$4.2~s, and
60$\pm$5.8~s, as shown in Figure~\ref{qpp3}~(c).

\begin{figure}
\includegraphics[width=\columnwidth]{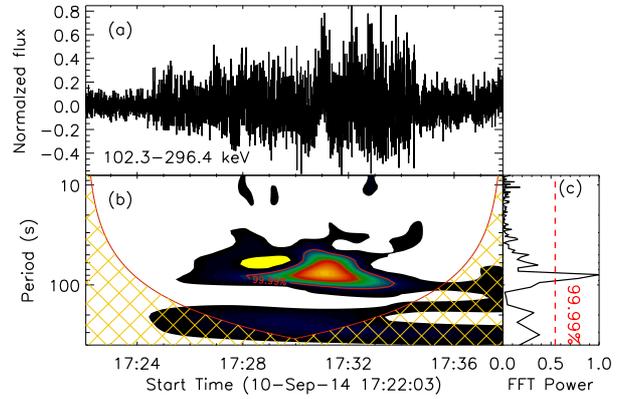}
\caption{Same as Figure~\ref{qpp1}, but at 102.3$-$296.4 keV.}
\label{qpp4}
\end{figure}

At higher energy channel (102.3$-$296.4~keV), we apply a wider
boxcar window of $\sim$100~s to smooth the light curve.
Figure~\ref{qpp4}~(a) shows the normalized rapidly varying signal.
The wavelet spectrum exhibits a period with a broad band between
17:28~UT and 17:33~UT, as shown in panel~(b). However, the FFT
spectrum only has one peak with a narrow width that greater than the
confidence level, which is around 80~s. Based on these facts, we
could obtain the dominant period within an error bar of 80$\pm$11~s
at higher energy channel. Even if using the short window timescale
($\sim$30~s), we could not find any periodic signal with shorter
period at higher energy channel. In other words, QPPs with shorter
period are only observed at lower energy channels, but fail to be
detected at higher energy channel.

\section{Discussions}
It is very interesting that the flare QPPs display multiple periods
in HXR emission during the impulsive phase of solar flare, and the
shorter periods tend to appear at lower energies of the X-ray
photons, while the longer periods appear at higher energies, with
some intersection of the periods at medium energies. The periods are
27~s and 37~s at lower photon energies (Figures~\ref{qpp1} and
\ref{qpp2}), and become around 80~s at higher photon energies
(Figure~\ref{qpp4}). Then the periods of 27~s, 46~s and 60~s are
found at medium photon energies (Figure~\ref{qpp3}). To our
knowledge, this is the first report that the QPPs display their
dominant periods become longer with the increasing of the photon
energies in HXR emission. Previous observations have reported the
various periods in multi-wavelengths. For example, \cite{Zimovets10}
report 16~s and 36~s QPPs at SXR and HXR channels ({\it RHESSI},
3$-$50~keV). \cite{Su12} detect flare QPPs with periods of
24$-$160~s at EUV band (AIA 171~{\AA}). \cite{Chowdhury15}
investigate QPPs during the decay phase of solar flare and find
double periods of $\sim$53~s and $\sim$72~s at SXR channels ({\it
RHESSI}, 3$-$25~keV). However, these observations just report the
multiple periods at one or several bands, but not indicate the
periods changing with the photon energies. Another interesting thing
is that the flare QPPs show their multiple periods appear at
different times during the impulsive phase of solar flare. This is
also a new result comparing with previous findings about
multi-period QPPs. For example, \cite{Reznikova11} find that the
periods become longer continuously in the nonthermal emission.
\cite{Hayes16} and \cite{Tian16} detect the different periods during
the flare impulsive and decay phases, respectively.

It is also interesting that the flare QPPs show the period ratio
within several values at HXR channels. At lower photon energies, the
period ratio is estimated to about 1.37 from the two dominant
periods; and at medium photon energy, the period ratios are around
2.22, 1.70 and 1.30, while at higher photon energy, there is only
one period value. On the other hand, the period ratio between
different photon energies can also be calculated. For example, the
period ratios are 2.96 and 2.16 between higher and lower photon
energies, and the period ratios of around 2.96, 1.74 and 1.33 can be
detected between higher and medium photon energies. The period
ratios between medium and lower photon energies are much more and
complex, since both of them display several dominant periods.
Similar as previous findings
\citep{Srivastava08,Inglis09,Su12,Tian16,Yang16}, all these period
ratios are deviating from two. This may be due to the expansion of
flare loop \citep{Verth08}, or the density stratification in solar
flare \citep{Andries05}.

The X1.6 flare on 2014 September 10 also exhibits QPPs with multiple
periods at various wavelengths, such as $\sim$240~s QPPs at HXR,
EUV/UV and radio emission \citep{Lit15,Li15,Dudik16}, $\sim$120~s
and $\sim$60~s QPPs at SXR/HXR and EUV/UV bands \citep{Ning17}. In
this paper, we obtain the multiple periods in HXR emission, i.e.,
27~s, 37~s, 46~s, 60~s and 80~s. The different periods of the flare
QPPs in pervious findings \citep{Li15,Ning17} are derived from the
different rapidly varying signals, which are dependent on the window
timescale. The window timescale of $\sim$240~s QPPs is 256~s
\citep{Li15}, and that of $\sim$60~s QPPs is 100~s \citep{Ning17}.
Both these QPPs can last up to around 18:00 UT at EUV/UV bands,
which is the decay phase of solar flare \citep{Li15,Ning17}.
Different from these previous observations, we find the shorter and
then the longer periods during the impulsive phase of this flare.

It is well known that the X-ray emission at lower and higher photon
energies can be produced by different mechanisms, i.e., by thermal
plasmas or by nonthermal electron beams. Usually, the thermal
component is extended up to 20~keV \citep{Ning08,Zhang16b}. And
sometimes, the nonthermal component can also be as low as about
10~keV \citep{Krucker02,Ning07,Ning08}. Therefore, the longer
periods at medium and higher photon energies are produced by
accelerated particles. And the shorter periods at lower photon
energies (12.0$-$27.3~keV) seem to be produced by thermal plasmas.
However, we could not find the shorter periods at 4.6$-$12.0~keV,
and the shorter periods can also be detected at 27.3$-$50.9~keV.
This is an indirect evidence that the QPPs are the features of
nonthermal emission. In a word, all the periods of flare QPPs in
X-ray emission during the impulsive phase of solar flare may be
produced by nonthermal electron beams. Meanwhile, fitting the energy
spectrum could proof this hypothesis, but it is out of scope of this
paper.

\section*{Acknowledgements}
The authors would like to thank the anonymous referee for his/her
valuable comments that improved the manuscript. We thank the teams
of {\it Fermi} and {\it GOES} for their open data use policy. This
study is supported by NSFC under grants 11603077, 11573072,
11333009, 973 program (2014CB744200), and Laboratory No.
2010DP173032. D.~Li is supported by the Youth Fund of Jiangsu No.
BK20161095. Q.~M.~Zhang is supported by the Surface Project of
Jiangsu No. BK20161618 and the Youth Innovation Promotion
Association CAS. The authors wish to thank the International Space
Science Institute in Beijing (ISSI-BJ) for supporting and hosting
the meetings of the International Team on ``Magnetohydrodynamic
Seismology of the Solar Corona in the Era of SDO/AIA", during which
the discussions leading to this publication were held.


\bsp
\end{document}